\documentclass[]{mn2e}

\usepackage{psfig,subfigure,mncite}

\newcommand{\bvec}[1]{\mbox{\boldmath ${#1}$}}

\begin{document}
 

\title[Starspot rotation rates]
{Stellar differential rotation from direct starspot tracking}

\author
[A. Collier~Cameron, J.-F. Donati and M. Semel]
{A. Collier Cameron$^1$
\thanks{E-mail: andrew.cameron@st-and.ac.uk},
J.-F. Donati$^2$,
M. Semel$^3$
\\
$^1$School of Physics and Astronomy, Univ.\ of St~Andrews, 
St~Andrews, 
Scotland KY16 9SS \\
$^2$Laboratoire d'Astrophysique, Observatiore Midi-Pyr\'{e}n\'{e}es,
Avenue E. Belin, F-31400 Toulouse, France
\\
$^3$DASoP, Observatoire de Paris-Meudon, F-92196 Meudon-Cedex, France
\\
} 

\date{Accepted 2001 November 2. Received 2001 October 31; in original form 
2001 September 20.}

\maketitle

\begin{abstract} 
On the Sun, the rotation periods of individual sunspots not only trace
the latitude dependence of the surface rotation rate, but also
provide clues as to the amount of subsurface fluid shear.  In this paper
we present the first measurements of stellar differential rotation
made by tracking the rotation of individual starspots with sizes
comparable to the largest sunspots.  To achieve this we re-analyse
four sequences of densely-sampled, high signal-to-noise echelle
spectra of AB Doradus spanning several stellar rotations in 1996
December.  Using spectral subtraction, least-squares deconvolution and
matched-filter analysis, we demonstrate that it is possible to measure
directly the velocity amplitudes and rotation periods of large numbers
of individual starspots at low to intermediate latitude.  We derive
values for the equatorial rotation rate and the magnitude of the
surface differential rotation, both of which are in excellent
agreement with those obtained by \scite{donati97doppler} from
cross-correlation of Doppler images derived a year earlier in 1995
December, and with a re-analysis of the 1996 data by the method of
\scite{donati2000rxj}.  The differences between the rotation rates of
individual spots and the fitted differential rotation law are
substantially greater than the observational errors.  The smaller
spots show a greater scatter about the mean relation than the larger
ones, which suggests that buffeting by turbulent supergranular flows
could be responsible.

\end{abstract}


\section{Introduction}

Observational studies of differential rotation on rapidly rotating
stars have recently been published by a number of authors. 
\scite{donati97doppler} and \scite{donati99doppler} applied a
latitude-by-latitude cross-correlation analysis to pairs of surface
images of the rapidly rotating K dwarf AB Doradus secured several days
apart.  They found a solar-like rate of surface shear, with the
equatorial region rotating faster than higher latitudes and a beat
period of about 110 days between the equatorial and high-latitude
rotation periods.  \scite{barnes2000pztel} obtained a similar result
for the young K dwarf PZ Tel.  \scite{donati2000rxj} used a parametric
imaging method to establish the differential rotation rate on the
pre-main sequence G star RXJ1508-4423 (= LQ Lup).  \scite{petit2000}
measured the differential rotation of the K subgiant primary of the RS
CVn binary HR 1099 by this same method.  \scite{vogt99hr1099} and
\scite{strassmeier2000hr1099} examined the evolving starspot
distribution on HR 1099 from long sequences of Doppler images, and
found evidence of a gradual poleward drift of large spot groups on
timescales of months to years.

Studies based on Doppler images suffer from a number of drawbacks for
this kind of work.  When the sampling of the line profile around the
stellar rotation cycle is sparse, the problem of mapping the stellar
surface-brightness distribution becomes an ill-conditioned inverse
problem, necessitating a regularised $\chi^{2}$ fitting approach.  The
regularising function imposes a penalty on the growth of fine-scale
structure in the stellar image.  This has the well-documented effect
of biassing the centroids of low-latitude features in the images away
from the stellar equator, and may even suppress them altogether.  It
it likely that differential rotation measurements based on these
methods may suffer some distortion as a result.

Several recent Doppler imaging studies of rapidly rotating young
main-sequence and pre-main sequence stars have, however, employed a
much denser sampling strategy.  With echelle spectrographs and
large-format CCD detectors it is possible to monitor the profiles of
thousands of lines simultaneously in continuous sequences of exposures
with a time resolution of a few minutes.  Signal-enhancement
techniques such as least-squares deconvolution (LSD:
\pcite{donati97zdi}) stack up the profiles of thousands of
photospheric lines to give composite line profiles with
signal-to-noise (S/N) ratios of several thousand.  

These densely-sampled series of composite line profiles delineate
clearly the trajectories of line-profile distortions of many small
spots.  They bear a superficial resemblance to time-series spectra of
rapidly rotating, early-type pulsating stars.  Many of the analytic
techniques that have been used to identify non-radial pulsation modes
in these objects operate directly on the data, without recourse to
large-scale inversion techniques
\cite{kennelly92,hao98,schrijvers99,jankov2000,telting2001}.  Since
the trajectories of starspot signatures in the trailed spectrogram of
a rotating star are sinusoidal, it should be possible to use analogous
forward-modelling techniques to identify individual spots and
determine their properties.

The purpose of this paper is to apply matched-filter analysis to the
problem of detecting and tracking individual unresolved starspots in
densely-sampled time sequences of high S:N line profiles.  Section 2
presents a re-analysis of archival echelle spectra of AB Doradus
observed in 1996 December, from which Doppler images have previously
been published.  Section 3 develops a matched-filter methodology based
on a similar technique used by Collier Cameron et al (1999, 2001)
\nocite{cameron99tauboo,cameron2001upsand} to search for faint
reflected-light signatures of extra-solar planets.  In section 4 this
technique is used to measure the rotation periods of individual spot
signatures present in each of these data sets, as a function of their
rotational velocity amplitudes.  In section 5 the results are compared
with those predicted by differential rotation models derived from
previous Doppler imaging studies.

\section{Data preparation}

The datasets employed here were secured using the SEMELPOL dual-fibre
polarimeter feed to the UCL echelle spectrograph (UCLES) at the 4.2-m
Anglo-Australian Telescope on 1996 December 23-29
\cite{donati99doppler}.  Readers are referred to the above paper for
the journal of observation and details of the instrument
configuration, and the initial extraction of the echelle spectra from 
the raw data frames.

The starspot signatures were isolated using a spectral subtraction and
least-squares deconvolution procedure.  The first task was to place
all the spectra on a common velocity scale.  This was achieved via the
least-squares deconvolution method of \scite{donati97zdi}.  The
essence of this technique is that the observed spectrum is treated as
the convolution of an ``average'' line profile and a list of delta
functions at the wavelengths of the known photospheric lines derived
from a Kurucz model.  The lines are weighted according to their
central depths, and the ``average'' profile is computed using the
method of least squares, such that it gives an optimal fit to the
spectrum when convolved with the line list.

\begin{figure}
	\psfig{figure=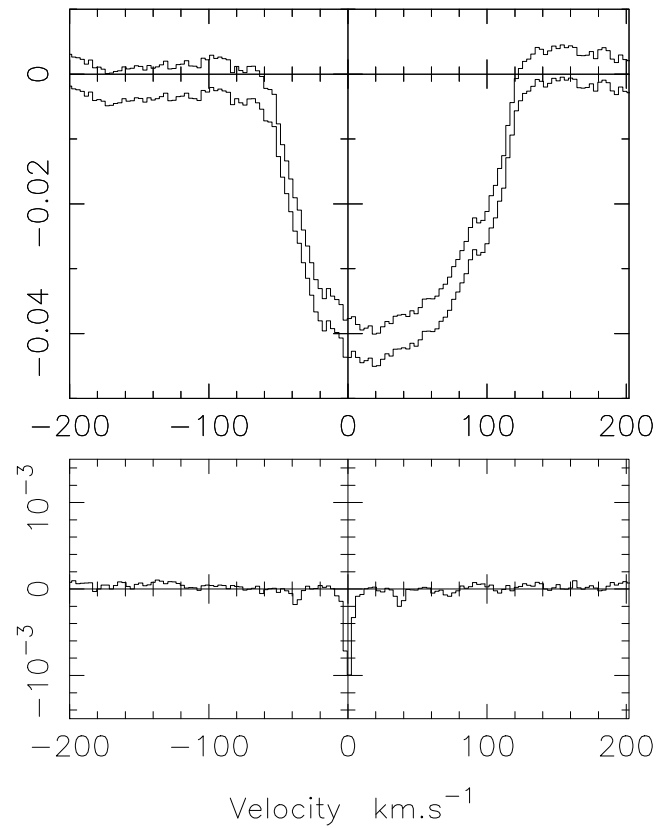,bbllx=27pt,bblly=7pt,bburx=310pt,bbury=370pt,width=8.6cm}
	\caption[]{The upper panel shows a sample least-squares
	deconvolved profile of AB Dor, computed using a list of stellar
	photospheric lines only (upper), and augmented with a list of
	telluric lines (lower).  The asymmetries and bumps in the profile
	are the signatures of starspots.  The difference between these two
	profiles (lower panel) shows a clear peak near zero velocity, 
	produced by telluric H$_{2}$O and O$_{2}$ lines.}
	\label{fig:tellurics}
\end{figure}

If the line list is augmented to include a list of known telluric
lines in the observed wavelength range, the deconvolved profile
develops a small, narrow pseudo-absorption bump near zero velocity. 
This feature becomes more apparent when we subtract the we subtract
the deconvolved profile computed using the stellar line list only, as
shown in Fig.~\ref{fig:tellurics}.  By fitting a gaussian to the
central peak of this residual profile, we are able to determine the
zero-point offset introduced by systematic errors in the wavelength
calibration (using Th-Ar arc spectra taken at the start and end of
each night) and by any instrumental flexure during the night.  This
offset is computed and applied to the zero-point of the velocity scale
for all subsequent deconvolutions, thereby ensuring that the
deconvolved spectra are accurately positioned in the observer's rest
frame.  

Telluric-line spectra derived from the rapidly-rotating B star HR 3084
were then scaled and subtracted from the individual echelle spectra. 
The individual spectra of AB Dor were then summed to create a time-averaged
spectrum for each night of observation.  This template
spectrum was then scaled and subtracted from each individual spectrum
in turn, using the spectral alignment technique described by
\scite{cameron2001upsand}.  

\begin{figure*}
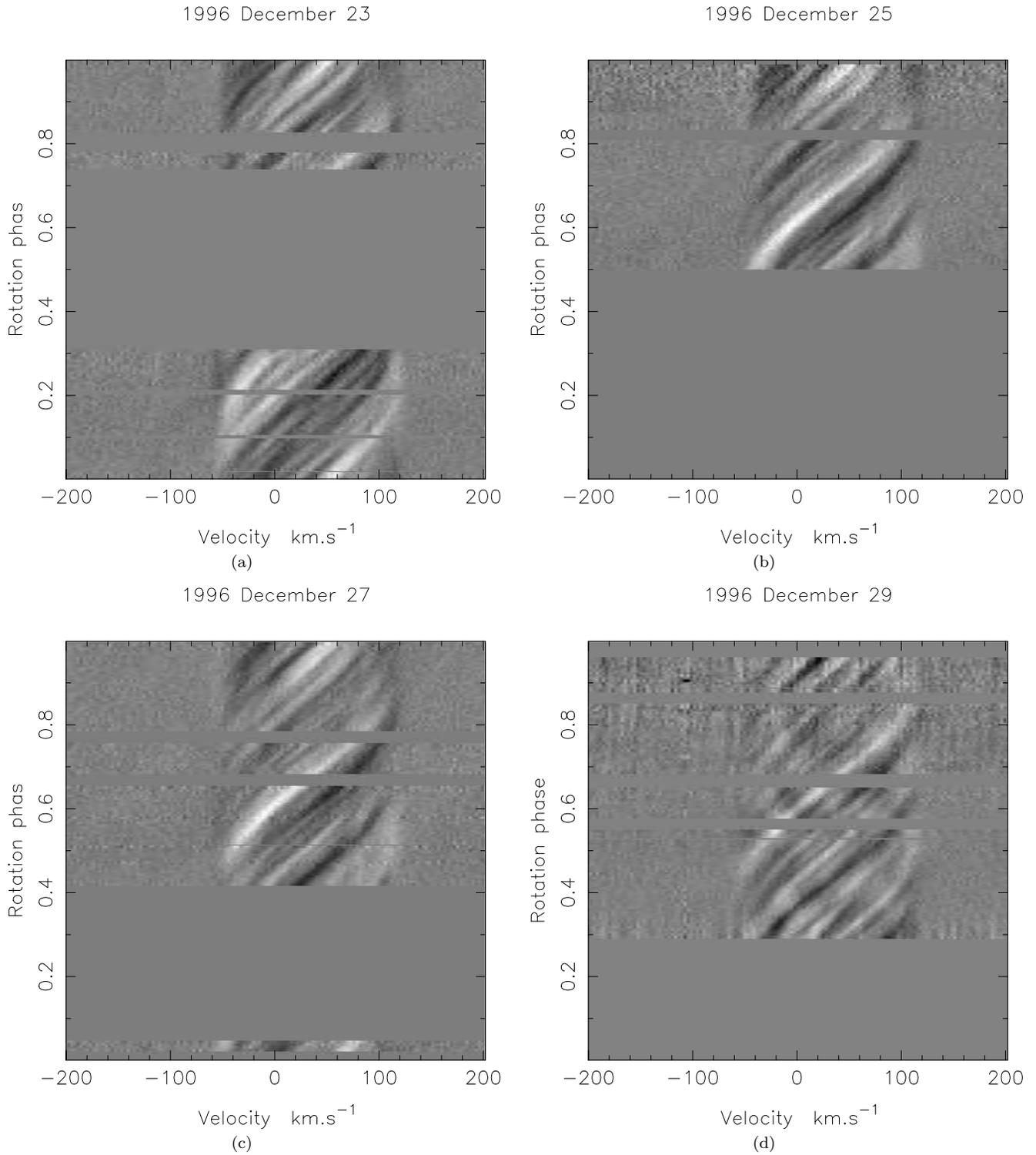

	\def\subfigtopskip{4pt}
	\def\subfigbottomskip{4pt}
	\def\subfigcapskip{2pt}
	\centering
	\begin{tabular}{ll}
    	\subfigure[]{
			\label{fig:decon23} 			
			\psfig{figure=fig02a.eps,bbllx=70pt,bblly=65pt,bburx=400pt,bbury=435pt,width=8.4cm}
			} &
		\subfigure[]{
			\label{fig:decon25} 
			\psfig{figure=fig02b.eps,bbllx=70pt,bblly=65pt,bburx=400pt,bbury=435pt,width=8.4cm}
			} \\
    	\subfigure[]{
			\label{fig:decon27} 
			\psfig{figure=fig02c.eps,bbllx=70pt,bblly=65pt,bburx=400pt,bbury=435pt,width=8.4cm}
			} &
		\subfigure[]{
			\label{fig:decon29} 
			\psfig{figure=fig02d.eps,bbllx=70pt,bblly=65pt,bburx=400pt,bbury=435pt,width=8.4cm}
			} \\
	\end{tabular} 
	\caption[]{Time-series
	spectra of deconvolved residual starspot signatures on the nights
	of 1996 December 23, 25, 27 and 29. The starspot signatures appear
	as bright pseudo-emission trails crossing the stellar rotation
	profile from negative
	to positive radial velocity. The rotation phases are
	computed according to the ephemeris of \scite{innis88},
	$\mbox{HJD}=2444296.575+0.51479E$.}
    \label{fig:dyndecon96}
\end{figure*}

This left a residual spectrum consisting mainly of random noise, on
which the residual line-profile distortions caused by stellar spots
were superimposed.  We deconvolved this residual spectrum with the
stellar line list, this time applying the heliocentric velocity
correction computed for the mid-time of the exposure.  This yielded
profiles from which the average line profile had been removed,
effectively isolating the starspot signatures and placing them
accurately on a heliocentric velocity scale.

Any residual telluric features produced by misalignment of the
telluric template and the target spectrum were removed at this stage
using principal-component analysis (PCA) of the night's ensemble of
spectra.  Removal of the first two principal components was sufficient
to eliminate the residual telluric lines without significantly
affecting the starspot trails.  A detailed description of the use of
PCA for removal of correlated noise features is given by
\scite{cameron2001upsand}.  In practice this only proved necessary for
the spectra observed on 1996 December 29.

The resulting time-series of deconvolved starspot residuals are shown
in Fig.~\ref{fig:dyndecon96}.  They show clearly the signatures of
individual starspots, following sinusoidal paths about the stellar
centre-of-mass velocity as they cross the visible hemisphere of the
star.  When the same range of rotation phases is observed on
successive nights, the same features are seen to recur at more or less
the same rotation phase.  As Donati et al (1997, 1999) have shown,
however, spots at high latitudes have slightly longer rotation periods
than spots near the equator.  High-latitude spots also have smaller
velocity amplitudes, since they lie closer to the stellar rotation
axis.  Since we are interested in measuring the rotation periods and
latitudes of individual spots in order to delineate the differential
rotation pattern, we need to develop an optimal method for determining
these two quantities directly from the trailed spectrograms.

\section{Matched-filter analysis}
\label{sec:matchfilt}

The spectral signature of an unresolved spot on the stellar surface is
closely approximated by a gaussian having the same full width at half
maximum intensity (FWHM) as the local specific intensity profile of
the spectral line concerned.

The gaussian profile varies in amplitude and radial velocity 
according to the expression:
$$
g(v,t | \phi,K,\Omega)=\frac{f(\gamma)}
{\Delta v_{\ell}\sqrt{\pi}}
\exp\left[
   -\left(
      \frac
	  {v - K \sin i \sin\psi - v_{0}}
	  {\Delta v_{\ell}}
   \right)^{2}
\right].
$$

Here $v_{0}$ is the radial velocity of the star's centre of mass.  A
spot located at latitude $\theta$ on a spherical star lies at a
distance $R_{\star}\cos\theta$ from the stellar rotation axis. 
If the stellar axis is inclined at an angle $i$ to the line of sight,
the spot's radial velocity will vary sinusoidally about $v_{0}$ with
angular frequency $\Omega(\theta)$ and amplitude $K$, where
\begin{equation}
K = \Omega(\theta)R_{\star}\cos\theta\sin i.
\label{eq:kvel}
\end{equation}
At any time $t$, a spot at position $(\phi,\theta)$ will be
displaced from the observer's meridian in the direction of rotation by
an angle
$$
\psi(\phi,\theta,t)=\Omega(\theta)(t-t_{0})+\phi.
$$
The stellar longitude $\phi$ increases in the direction of
rotation, and is measured from the observer's meridian on the stellar
surface at an arbitrary time $t_{0}$. The fiducial epoch $t_{0}$ is 
chosen to be near the middle of the sequence of observations, to 
eliminate spurious correlations between $\phi$ and $\Omega$.

While the spot is on the visible hemisphere its apparent area will be
foreshortened by a factor 
$$
\cos\gamma = \cos i \sin\theta + \sin i \cos\theta \cos\psi.
$$ 
Here $\gamma$ is the angular separation of the spot and the observer
subtended at the centre of the star. 

The flux blocked by the spot is also subject to limb darkening,
which is assumed to have a linear dependence on $\cos\gamma$ with a
limb-darkening coefficient $u=0.77$.  This value, from the tabulation
of \scite{diazcordoves95}, is appropriate for an early-K photosphere
at the depth- and inverse-variance weighted mean wavelength
$\hat{\lambda}=5475$\AA\ of the spectral lines used in the
deconvolution.

The overall reduction in signal strength for a spot at foreshortening
angle $\gamma$ relative to its strength when viewed directly from
above ($\gamma = 0$) is
$$
f(\gamma) = \cos\gamma(1-u+u\cos\gamma).
$$

The gaussian width parameter $\Delta v_{\ell}$ is related to the line
FWHM by
$$
\Delta v_{\ell} = \frac{\mbox{FWHM}}{2\sqrt{\ln 2}}.
$$
The spectrum of the slowly rotating K dwarf Gl 176.3 has been used in
several previous Doppler imaging studies of AB Dor as a spectral
template whose line profiles are representative of the local line
profiles on AB Dor in the absence of rotational broadening.  A
gaussian fit to the least-squares deconvolved profile of Gl 176.3
yielded $\Delta v_{\ell} = 5.53$ km s$^{-1}$.

The matched filter $g(v,t | \phi,K,\Omega)$ can be computed on a
two-dimensional map with the same dimensions as the observed
time-series spectrum, and fitted to the data via an optimal scaling. 
The scale factor $W$ by which $g$ is multiplied to give the best fit
to an individual spot trail, is the equivalent width of the bump in
the line profile that would be produced by the spot if it were located
at the centre of the disc.  It is proportional to the area of the
spot, and related to the intensity ratio of the local continuum in the
photosphere and the spot, the ratio of the line equivalent widths in
the photosphere and the spot, the overall spot coverage, and the
limb-darkening coefficient.

\subsection{Second-derivative profiles}

\begin{figure}
	\psfig{figure=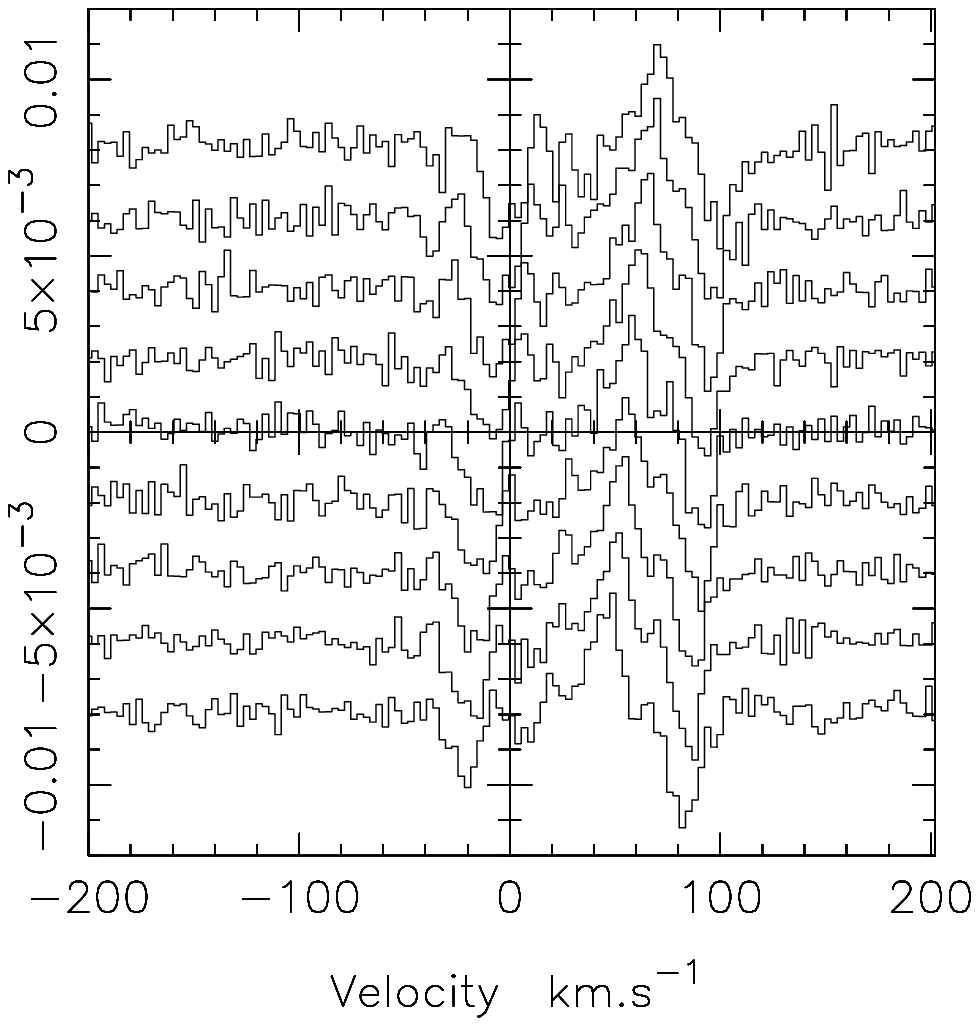,bbllx=27pt,bblly=7pt,bburx=310pt,bbury=300pt,width=8.6cm}
	\caption[]{Sample deconvolved profiles from the trailed
	spectrogram for 1996 December 25 between phases 0.6 and 0.7 show
	closely-spaced peaks caused by marginally-resolved spots,
	separated by as little as 5 pixels. The residual intensity units 
	are given as a fraction of the mean continuum intensity.}
	\label{fig:minsep}
\end{figure}

The matched-filter fitting procedure described above suffers from the
disadvantage that although the residual starspot signatures are
approximately gaussian, they appear superimposed on a sloping,
non-zero background that changes slowly from one spectrum to the next
(Fig.~\ref{fig:minsep}).  This means that $W$ cannot be determined
using a simple gaussian fit.

Instead, we compute a numerical second derivative to pick out the 
``ridge line'' traced out by a spot in the trailed spectrogram.  
The second derivative of each spectrum $\bvec{s}$
in the observed time-series is defined as:
$$
x_{i,j} = s_{i-2,j} - 2s_{i,j} + s_{i+2,j}.
$$
The choice of a two-pixel offset is a compromise between computing the
derivatives over a small velocity range to avoid loss of resolution,
and maximising the baseline over which the derivatives are computed to
reduce noise.  The smallest resolvable separation between neighbouring
spot features appearing in a single line profile is roughly 5 pixels
(Fig.~\ref{fig:minsep}), which suggests that a two-pixel offset is
optimal in this respect.

\begin{figure*}
	\def\subfigtopskip{4pt}
	\def\subfigbottomskip{4pt}
	\def\subfigcapskip{2pt}
	\centering
	\begin{tabular}{ll}
    	\subfigure[]{
			\label{fig:phch5125} 						
			\psfig{figure=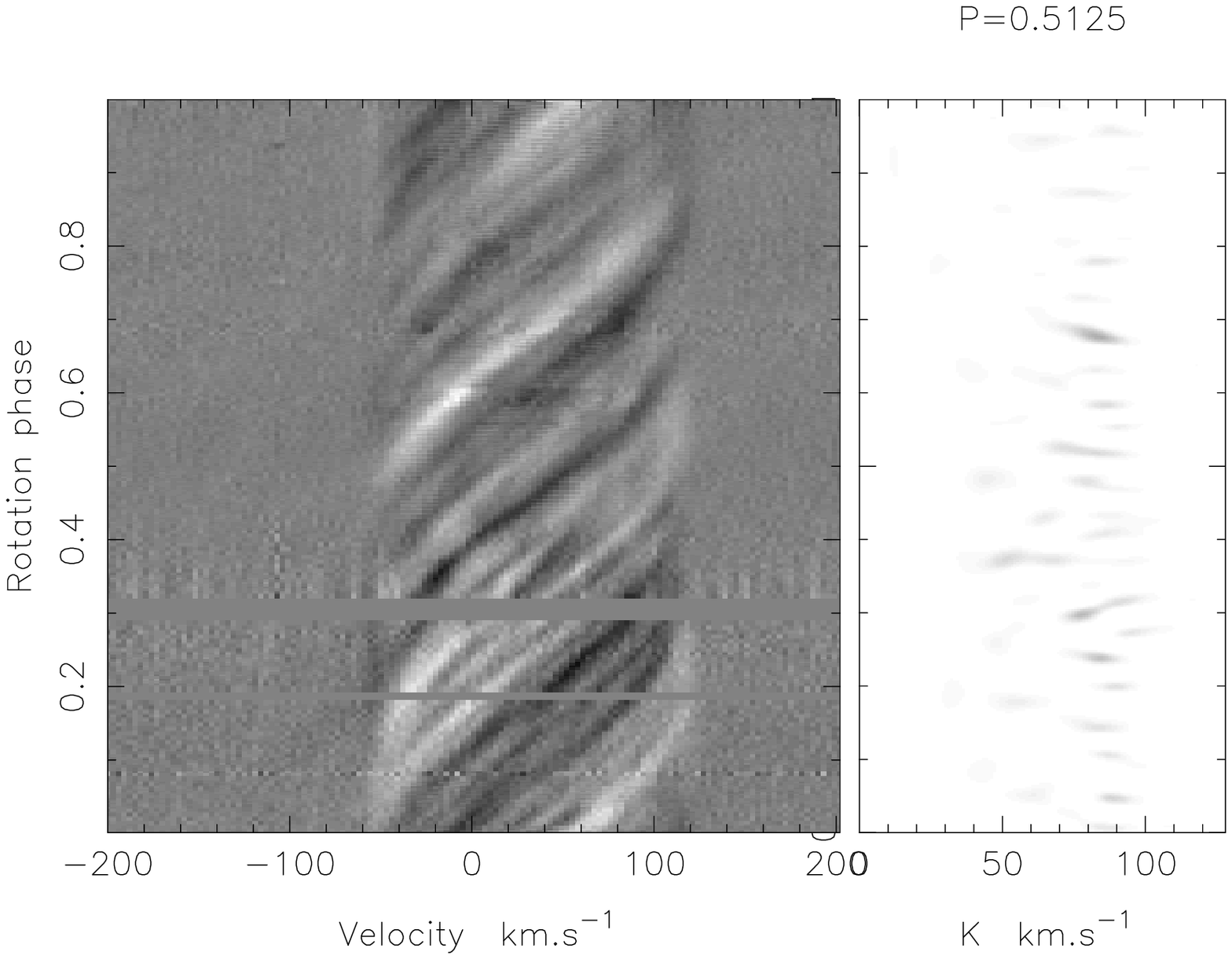,bbllx=70pt,bblly=65pt,bburx=549pt,bbury=435pt,width=8.4cm}
			} &
		\subfigure[]{
			\label{fig:phch5135} 			
			\psfig{figure=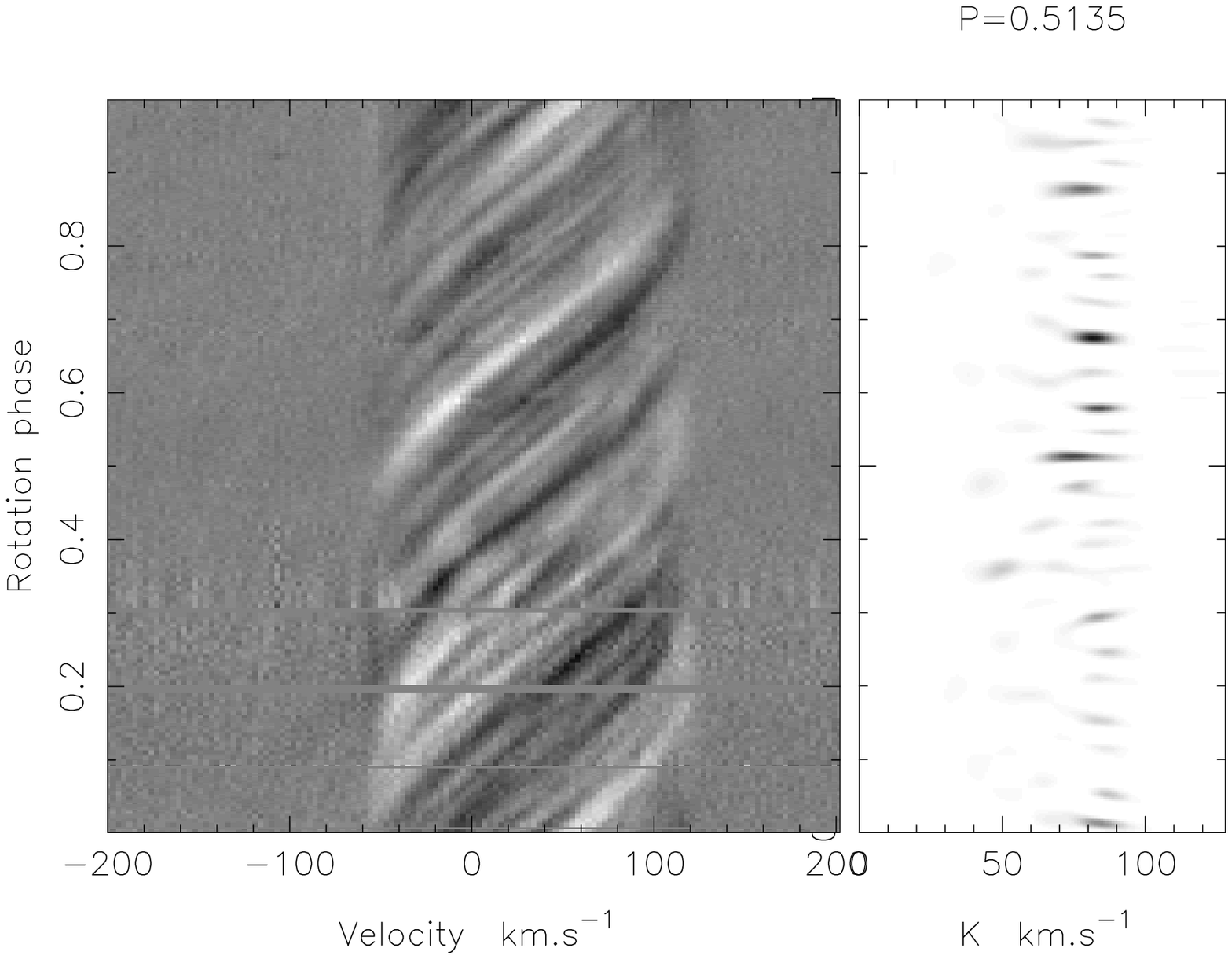,bbllx=70pt,bblly=65pt,bburx=549pt,bbury=435pt,width=8.4cm}
			} \\
    	\subfigure[]{
			\label{fig:phch5145} 			
			\psfig{figure=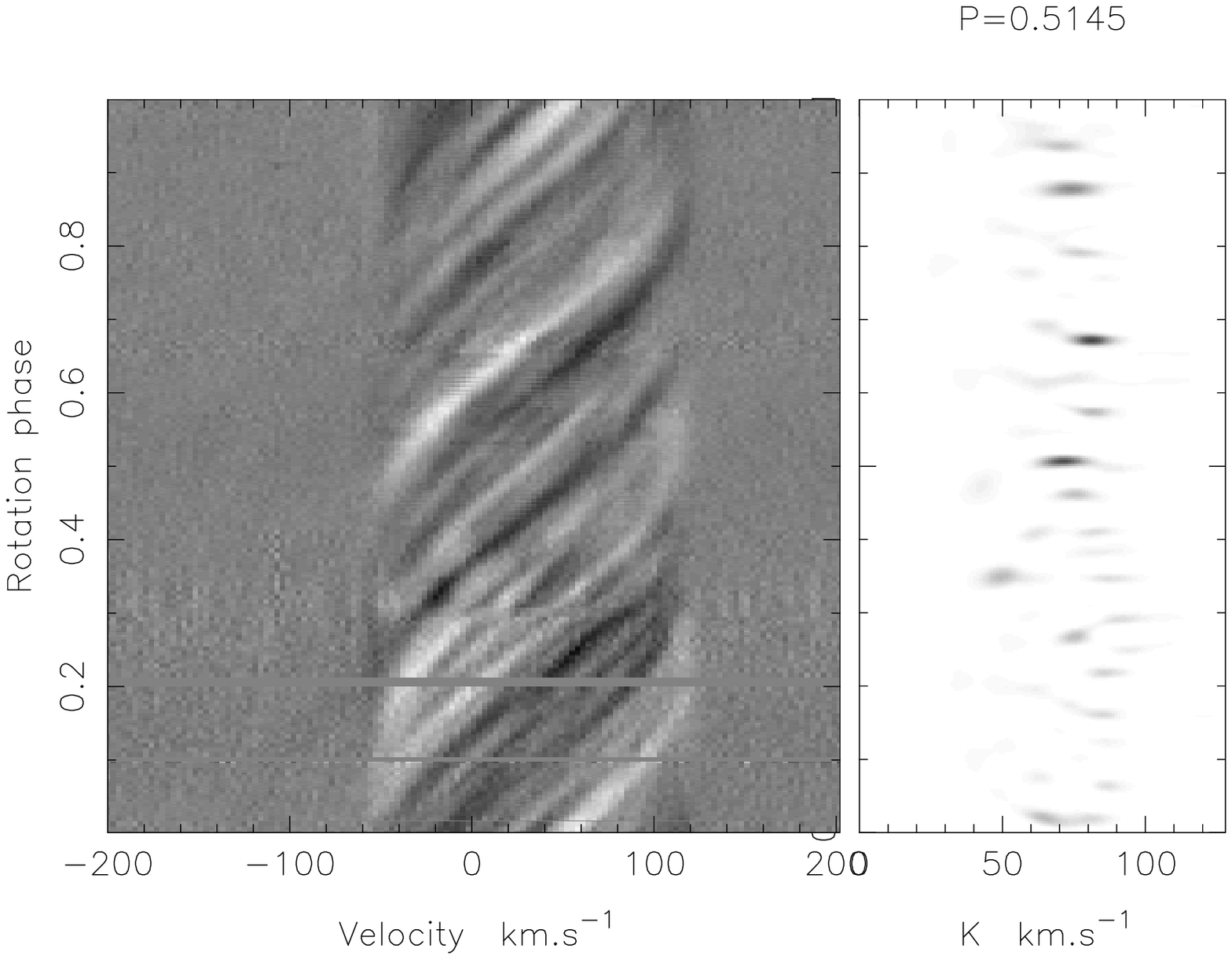,bbllx=70pt,bblly=65pt,bburx=549pt,bbury=435pt,width=8.4cm}
			} &
		\subfigure[]{
			\label{fig:phch5155} 			
			\psfig{figure=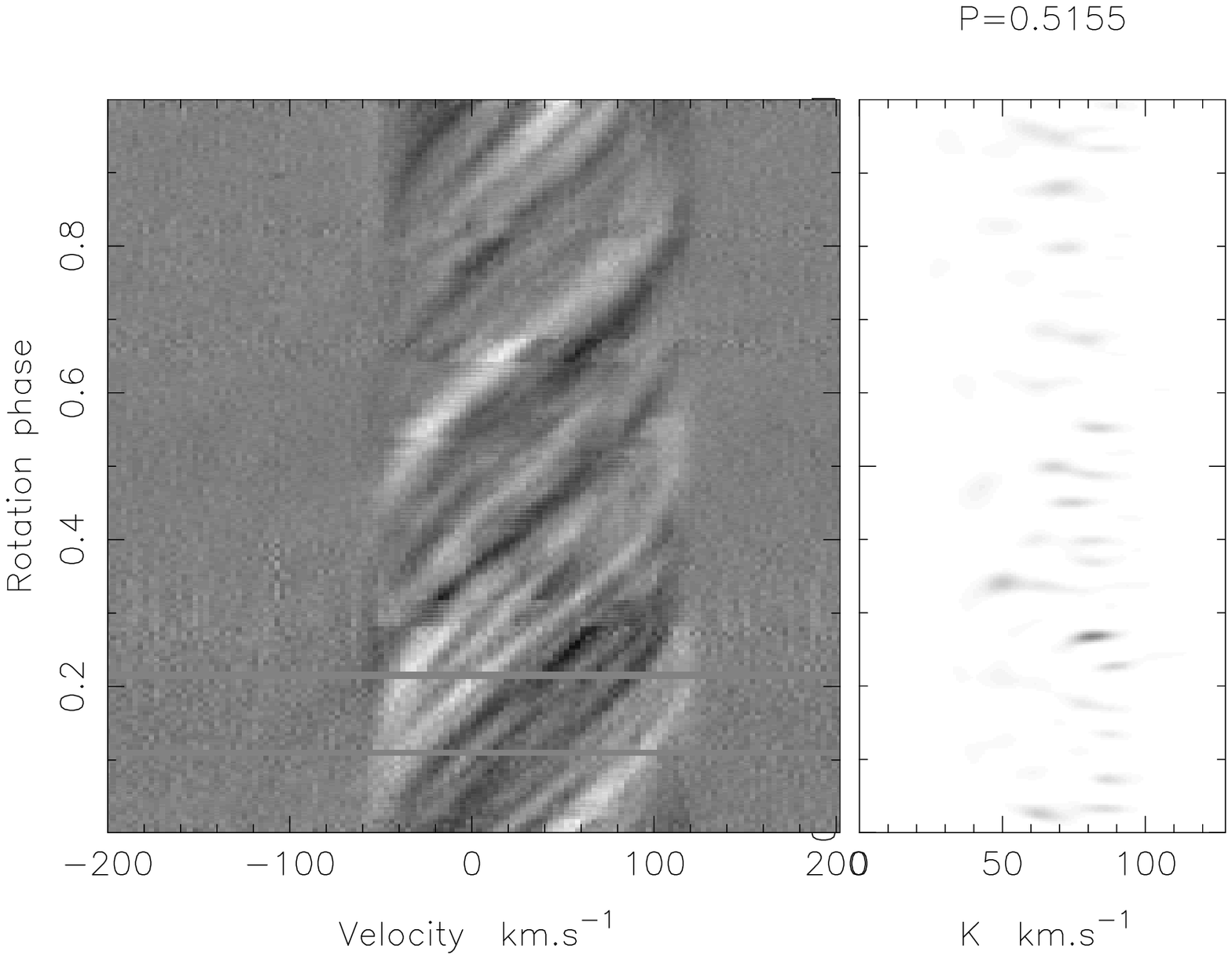,bbllx=70pt,bblly=65pt,bburx=549pt,bbury=435pt,width=8.4cm}
			} \\
	\end{tabular} 
	\caption[]{The right-hand subpanels show $\chi^2$ as a function
	of rotational velocity amplitude $K$ and phase, for four values of
	the rotation period: (a) 0.5125 d; (b) 0.5135 d; (c) 0.5145 d; (d)
	0.5155 d.  The corresponding trailed spectrograms in the left-hand
	subpanels show all four nights' data merged and phased using these
	rotation periods.  A dark, oval $\chi^2$ peak in the right-hand
	panel corresponds to a bright starspot trail (white) crossing the centre
	of the line profile in the left subpanel at the same rotation
	phase.  The velocity amplitude $K$ in the right-hand panel is
	proportional to the radial acceleration of the spot signature as
	it crossed the centre of the profile.  The definition of
	fine-scale features in the trailed spectrum is very sensitive to
	the period on which the data are folded.}
	\label{fig:cubeplanes}
\end{figure*}

We treat the matched filter in the same way:
$$
p_{i,j} = f_{i-2,j} - 2f_{i,j} + f_{i+2,j}.
$$
Here $f_{i,j}$ is the contribution of the gaussian matched filter to
the $i$th pixel of the $j$th profile in the time-series, which spans
the velocity range from $v_{i-1/2}$ to $v_{i+1/2}$:
$$
f_{i,j}=\int_{v_{i-1/2}}^{v_{i+1/2}} g(v_{i},t_{j} | \phi,K,\Omega)\, dv.
$$
The scale factor $\hat{W}$ is then computed from these
second-derivative maps via the optimal scaling:
$$
\hat{W} = \frac{\sum_{i,j}x_{i,j}p_{i,j}/\mbox{Var}(x_{i,j})}
{\sum_{i,j}p^{2}_{i,j}/\mbox{Var}(x_{i,j})}.
$$
Here $\sigma^{2}_{i,j}$ is the variance associated with the data point 
$s_{i,j}$, so
$$
\mbox{Var}(x_{i,j})=\sigma^{2}_{i-2,j} + 4\sigma^{2}_{i,j} + \sigma^{2}_{i+2,j}.
$$
Note that the scale factor $\hat{W}$ is the equivalent width
in velocity units of the starspot
bump that would be seen in the deconvolved profile if the spot were
observed at the centre of the stellar disc, and so is expressed in
km~s$^{-1}$.

\section{Results}

\subsection{Spot parameter estimation}

The optimal fitting procedure determines both $\hat{W}$ 
and the associated badness-of-fit statistic
$$
\chi^{2}=\sum_{i,j}\frac{(x_{i,j}-\hat{W}p_{i,j})^{2}}{6\sigma^{2}_{i,j}}
$$
for a given set of matched-filter parameters $(\phi,K,\Omega)$. To 
identify spot signatures and measure their parameters, we construct a 
sequence of maps on a $(\phi,K)$ grid, for a range of rotation 
periods $P=2\pi/\Omega$. 

The sequence of maps is stacked to form a data cube,
$\hat{W}(\phi,K,\Omega)$ or $\Delta\chi^2(\phi,K,\Omega)$ which is then
searched for local maxima in all three dimensions.  Inevitably some of
the weaker local maxima are spurious noise features, so it is
necessary to impose a cutoff on $\hat{W}$ (or better, on
$\Delta\chi^{2}$) below which candidate spots are rejected as being probable
noise features.  An appropriate cutoff is found by computing the
$(\phi,K)$ maps over a range of $K$ values somewhat greater than
the stellar $v\sin i$, and tuning the cutoff to reject all features
with $K > v\sin i$.

Note also from Fig.~\ref{fig:dyndecon96} that good repeated phase
coverage was obtained on at least 3 of the 4 nights only for features
crossing the centre of the line profile in the phase range from 0.50
to 1.00.  The search was therefore restricted to this phase range.

\begin{table}
	\caption[]{Parameters derived from matched-filter analysis for the
	24 candidate starspot signatures in phase range 0.50 to 1.00 on
	1996 December 23, 25, 27 and 29.  The six asterisked features
	are aliases.}
	\begin{tabular}{cccl}		
		Phase & $K$ & $P$ & $\hat{W}$  \\
		      &(km s$^{-1}$) & (days) & (km s$^{-1}$) \\
		      &                   &            &                 \\
$0.5099	\pm 0.0001$ & $	73.29\pm 0.07$ & $0.513964\pm 0.000023$ & 0.00655 \\
$0.5471	\pm 0.0002$ & $	87.32\pm 0.34$ & $0.513299\pm 0.000037$ & 0.00297 \\
$0.5504	\pm 0.0007$ & $	59.03\pm 0.60$ & $0.513966\pm 0.000088$ & 0.00171 \\
$0.5514	\pm 0.0003$ & $	83.41\pm 0.29$ & $0.515664\pm 0.000036$ & 0.00298* \\
$0.5788	\pm 0.0001$ & $	83.47\pm 0.04$ & $0.513597\pm 0.000020$ & 0.00596 \\
$0.5878	\pm 0.0004$ & $	86.79\pm 0.19$ & $0.516632\pm 0.000063$ & 0.00176* \\
$0.6258	\pm 0.0015$ & $	38.10\pm 0.47$ & $0.513507\pm 0.000371$ & 0.00155* \\
$0.6284	\pm 0.0003$ & $	80.67\pm 0.18$ & $0.513437\pm 0.000055$ & 0.00267 \\
$0.6401	\pm 0.0003$ & $	82.32\pm 0.09$ & $0.516597\pm 0.000052$ & 0.00196* \\
$0.6737	\pm 0.0001$ & $	81.48\pm 0.12$ & $0.513839\pm 0.000019$ & 0.00715 \\
$0.6919	\pm 0.0005$ & $	64.77\pm 0.36$ & $0.514308\pm 0.000083$ & 0.00244 \\
$0.7252	\pm 0.0003$ & $	81.30\pm 0.17$ & $0.513334\pm 0.000047$ & 0.00254 \\
$0.7594	\pm 0.0002$ & $	86.06\pm 0.10$ & $0.513738\pm 0.000037$ & 0.00297 \\
$0.7878	\pm 0.0002$ & $	81.04\pm 0.09$ & $0.513662\pm 0.000026$ & 0.00413 \\
$0.8129	\pm 0.0006$ & $	67.23\pm 0.66$ & $0.513837\pm 0.000065$ & 0.00181 \\
$0.8134	\pm 0.0004$ & $	91.01\pm 0.16$ & $0.513032\pm 0.000066$ & 0.00158 \\
$0.8247	\pm 0.0007$ & $	51.14\pm 0.14$ & $0.515061\pm 0.000092$ & 0.00188 \\
$0.8788	\pm 0.0002$ & $	76.77\pm 0.14$ & $0.513847\pm 0.000026$ & 0.00547 \\
$0.9139	\pm 0.0002$ & $	89.35\pm 0.11$ & $0.513543\pm 0.000035$ & 0.00275 \\
$0.9327	\pm 0.0003$ & $	85.78\pm 0.46$ & $0.515426\pm 0.000057$ & 0.00240* \\
$0.9365	\pm 0.0003$ & $	70.46\pm 0.25$ & $0.514310\pm 0.000037$ & 0.00369 \\
$0.9580	\pm 0.0004$ & $	81.67\pm 0.12$ & $0.515077\pm 0.000055$ & 0.00201* \\
$0.9580	\pm 0.0007$ & $	60.90\pm 0.52$ & $0.515384\pm 0.000081$ & 0.00254 \\
$0.9687	\pm 0.0003$ & $	85.76\pm 0.06$ & $0.513512\pm 0.000033$ & 0.00325 \\
	\end{tabular}
	\label{tab:spotpar}
\end{table}

This yielded the list of candidate spots in Table~\ref{tab:spotpar}. 
The parameter values are listed together with their formal $1\sigma$
errors, derived using
$$
\mbox{Var}(K) = \frac {2}{\partial^{2}\chi^{2}/\partial K^{2}}
$$
and similarly for $\phi$ and $P$.  The $1\sigma$ errors on the spot velocity
amplitude and period are consistent with the values expected from
simple considerations: if the radial velocity of a spot bump can be
determined to a precision of order 0.1 pixel (300 m~s$^{-1}$), the
uncertainty in the time at which it crosses the observer's meridian is
of order 20s.  Over a 6-day (12-rotation) baseline, this gives an
uncertainty of about 0.00003 day in the period determination. 
Determining the radial acceleration -- and hence $K$ -- involves
measuring the change in spot velocity over the course of roughly
one-sixth of the rotation cycle, which again involves velocity and
timing measurements that are uncertain at the 0.1 km~s$^{-1}$ and 20s
levels respectively.  The resulting uncertainty in $K$ ought therefore
to be of order 0.1 to 0.2 km~s$^{-1}$ for a well-observed spot.

We explored the effects of possible errors in the limb-darkening
coefficient by repeating the matched-filter analysis using a more
solar-like $u=0.6$ rather than $u=0.77$.  Given that the time-averaged
spectrum is subtracted prior to deconvolution, the principal effect of
decreasing the limb-darkening coefficient is to decrease the contrast
between the strength of a spot signature viewed at the centre of the
disc and near the stellar limb.  As expected, we found that the lower
value of $u$ yielded a slightly poorer fit to the data.  The velocity
amplitudes and rotations periods of individual spots were found,
however, to be nearly identical (to within the 1-$\sigma$ error
limits) for both values of $u$.  This confirms that the results are
insensitive to any error in the adopted value of the limb-darkening
coefficient.

As was mentioned in Section~\ref{sec:matchfilt} above, the equivalent 
width of a spot bump is related to the area of the spot.
To a very good approximation, the equivalent width of an absorption 
line in the specific intensity profile from any part of the stellar disc 
is the same as the equivalent width of the rotationally-broadened line 
profile considered as a whole. The ratio of the equivalent width of a 
small starspot bump at disc centre to that of the entire line profile is 
therefore just the ratio of the flux ``missing'' in the spot ($F_{spot} 
\simeq 
\pi I_0 (R_{spot}/d)^{2}$) to the total flux from the remaining limb-darkened
photosphere,
$F_{phot} = \pi I_0 (R_{\star}/d)^2  (1-u/3)-F_{spot}$. Here $I_{0}$ is 
the specific intensity of the photosphere at disc centre, 
$R_{spot}$ and $R_{\star}$ are the radii of the spot and the 
star respectively, and $d$ is the distance to the star. Hence
$$
\frac{W_{spot}}
     {W_{tot}}
\simeq
\frac{1}{1-u/3}
\left(\frac{R_{spot}}{R_{\star}}\right)^{2}
$$
for a small, isolated spot viewed at the centre of the stellar disc.

In reality the spots are neither isolated nor completely dark.  At the
red wavelengths observed, the continuum surface brightness of the
spots is expected to be about 0.25 to 0.3 times the photospheric
value.  The bump amplitude for a spot of given area is thus expected
to be diminished by this amount relative to the bump amplitude for a
completely dark spot.  However, the fractional coverage of spots on AB
Dor is probably of order 30 percent, which serves to increase the loss
of light from an individual spot as a fraction of the total amount of
``clean'' photosphere.  Overall, the effects of finite spot surface
brightness and a high spot filling factor will tend to cancel each
other out.

The spots listed in Table~\ref{tab:spotpar} have equivalent widths
ranging from $W=0.0016$ to $W=0.0071$ km s$^{-1}$.  Since the
equivalent width of the deconvolved stellar profile is approximately
4.5 km~s$^{-1}$ (cf.  Fig.~\ref{fig:tellurics}), the inferred 
fractional areas $(R_{spot}/R_{\star})^{2}$ of
individual spots range from 0.00026 to 0.0012 of the visible
hemisphere.  AB Dor has a radius similar to the Sun, so if they are
considered as single, circular spots, their radii range from 11300 km
up to 24000 km.  The smallest detectable spots on AB Dor are thus
comparable with the largest observed sunspots, whose umbral radii
occasionally exceed 10000 km.

\subsection{Aliasing between neighbouring spots}

\begin{figure}	
	\psfig{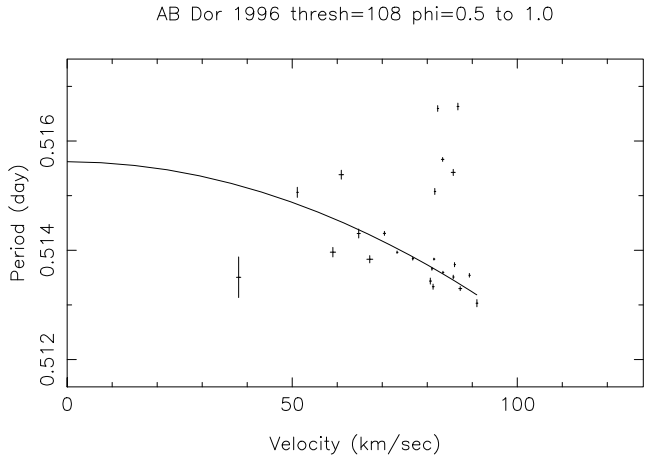}
	\caption[]{Rotation period $P$ versus rotational velocity
	amplitude $K$ for candidate starspots in phase range 0.50 to
	1.00.  Note that the spot properties fall into two main
	groups.  The main cluster forms a sloping sequence delineating
	the stellar surface differential rotation.  The group at top
	right are produced by aliasing between closely-spaced spots
	observed near the beginning and end of the run.  The solid
	curve shows the fitted differential rotation law.}
	\label{fig:diffrot}
\end{figure}

The relationship between the velocity amplitudes $K$ and rotation
periods $P$ of the candidate spots is plotted in
Fig.~\ref{fig:diffrot}.  Although many of the spots lie on a sloping
sequence delineating the stellar surface differential rotation
pattern, a second cluster of points with high values of $K$ and
anomalously long periods is also present.

The upper cluster of points appears to result from aliasing between
closely-spaced pairs of low-latitude spots.  The top panel of
Fig.~\ref{fig:cubeplanes} shows numerous low-latitude spots, with a
typical separation in phase of between 0.02 and 0.04 rotations.  The
candidates with $K>80$ km~s$^{-1}$ and $P>0.515$ day are invariably
sandwiched between a pair of candidates at similar latitudes, with
periods of order 0.5135 day with a phase difference of 0.02 to 0.03,
observed 4 to 6 days apart.  Aliases will occur when the data set is
phased at periods which cause a genuine spot signature seen near
the start of the run to appear superimposed on that of a neighbouring
spot at similar latitude observed near the end of the run.  Aliases
will thus occur at periods given by
$$
P_{\mbox{alias}}=\left(\frac{1}{P_{\mbox{true}}}\pm
\frac{1}{P_{\mbox{beat}}}\right)^{-1}.
$$
A phase shift of 0.03 rotations over a 4-day interval 
corresponds to a beat period of 133 days. Hence we expect a pair of 
spots with $P=0.5135$ days separated by 0.03 in phase to produce
aliases at periods of 0.5155 and 0.5115 days. Pairs of spots with
separations of 0.02 to 0.04 rotations, observed 4 to 6 days apart,
can thus produce aliases with periods anywhere between 0.515 and 
0.5165 days. This corresponds closely to the period range in which 
the aliases are seen in Fig.~\ref{fig:diffrot}. The weak candidate
spot at $K\simeq  38$ km~s$^{-1}$ and $P\simeq 0.5134$ days appears
to be another type of alias, arising from crosstalk between the
blue-shifted and red-shifted ends of two distinct starspot trails
from low-latitude features at phases 0.58 and 0.67.

\subsection{Fitting a differential rotation law}

\scite{donati97doppler} found that a differential rotation law of the form
\begin{equation}
\Omega(\theta) = \Omega_{eq} - \Omega_{b}\sin^{2}\theta
\label{eq:omega}
\end{equation}
with $\Omega_{eq}=12.2434$ rad day$^{-1}$ and $\Omega_{b}=0.0564$ rad
day$^{-1}$ gave a good description of the latitude dependence of the
surface rotation rate of the magnetic features and dark starspots on
AB Dor in 1995 December.  This corresponds to an equatorial rotation
period $P_{\mbox{equator}}=2\pi/\Omega_{eq}=0.5132$ days, and an
equator-pole lap time $P_{\mbox{beat}}=2\pi/\Omega_{b}=111$ days.

Combining eqs. (\ref{eq:kvel}) and (\ref{eq:omega}) gives 
\begin{equation}
K = v\sin i\cos\theta(1-\frac{\Omega_{b}}{\Omega_{eq}}\sin^{2}\theta),
\label{eq:ktheta}
\end{equation}
which can be solved iteratively to obtain the stellar latitude 
$\theta$ at which a spot will have a projected rotational velocity 
amplitude $K$, given the values of  $v\sin i$, $\Omega_{eq}$ and 
$\Omega_{b}$. The rotation frequency $\Omega$ of the spot (and hence
its period) follows from eq. (\ref{eq:omega}).

The equatorial rotation speed is known accurately from many previous
Doppler imaging studies to be $v\sin i=91\pm 1$ km~s$^{-1}$.  We can
therefore use eqs.  (\ref{eq:omega}) and (\ref{eq:ktheta}) to
determine the values of $\Omega_{eq}$ and $\Omega_{b}$ that give the
best (least-squares) fit to the data.

\begin{figure}
	\psfig{file=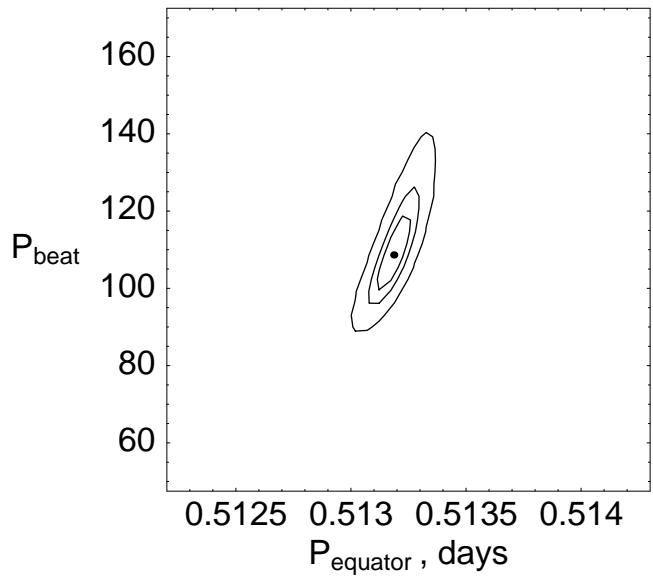,width=8.6cm}
	\caption[]{Contours of $\chi^{2}$ as a function of equatorial 
	rotation period and equator-pole lap time. Both periods are 
	expressed in days. The inner contour is the locus of values 
	$\chi^{2}=\chi^{2}_{\mbox{min}}+1.0$, whose extremities
	in $P_{\mbox{equator}}$ and $P_{\mbox{beat}}$ give the
	one-dimensional $1\sigma$ error bars on the two periods. The outer 
	contours, at $\chi^{2}=\chi^{2}_{\mbox{min}}+2.3$ and 
	$\chi^{2}=\chi^{2}_{\mbox{min}}+6.2$, contain 68.3\%\ and 95.4\%\ of 
	the joint probability respectively.}
	\label{fig:contours}
\end{figure}

Fig.~\ref{fig:contours} shows contours of $\chi^{2}$ as a function of
$P_{\mbox{equator}}$ and $P_{\mbox{beat}}$.  Note that, since the
scatter in the period residuals is much greater than the precision of
the measurements, an additional source of variance must be added to
the errors associated with the periods listed in Table
\ref{tab:spotpar}.  As discussed in Section~\ref{sec:resid} below, the
additional variance appears to be intrinsic to the star, and is
equivalent to an additional RMS uncertainty of $\pm 0.00017$ d in the
period determination for each spot.  Adding this error in quadrature
to the measurement errors reduced the $\chi^{2}$ for the fit such that
its minimum value was of order the number of degrees of freedom (in
this case, 18 data points minus 2 fitted parameters.)

This fit yields $P_{\mbox{equator}}=0.51319\pm 0.00007$ days, and
$P_{\mbox{beat}}=109\pm 9$ days.  The parameters $P_{\mbox{equator}}$
and $P_{\mbox{beat}}$ are strongly correlated.  The differential
rotation parameters for the 1995 December dataset, obtained by
\scite{donati97doppler}, lie within the inner contour, close to the
major axis of the error ellipse, indicating excellent agreement
between the two methods.  A re-analysis of the 1996 December data
using the sheared-image method of \scite{donati2000rxj} gives
$P_{\mbox{equator}}=0.51332\pm 0.00012$ days, and
$P_{\mbox{beat}}=115\pm 15$ days, again in good agreement.

\subsection{Residuals and spot size}
\label{sec:resid}

\begin{figure}
    	\psfig{file=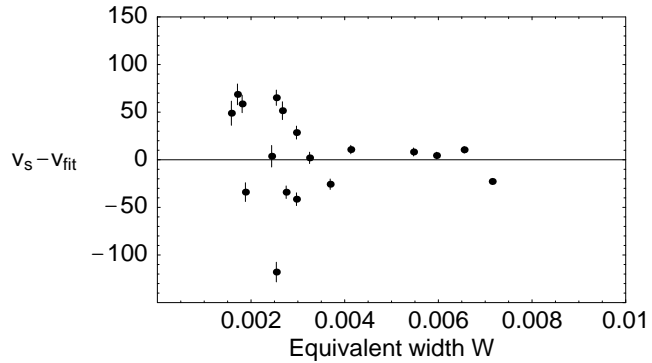,width=8.6cm}
	\caption[]{Residual rotational velocities 
        (expressed in m~s$^{-1}$) from the fitted differential 
        rotation law, plotted as a
	function of spot equivalent width $W$.
        Note the increased scatter in the residuals
	for the smaller spots.}
	\label{fig:resid}
\end{figure}

The scatter about the fit to the data shown in Fig.~\ref{fig:diffrot}
is considerably greater than the sizes of the errors on the data
points.  The measured periods of some of the best-observed spots
deviate from the mean relation by as much as 0.005 day, giving an
accumulated error of order an hour over the 6-day span of the data. 
This is too great to be ascribed to any plausible source of systematic
error, such as irregularly-shaped spot groups spanning a range of
latitudes.  We are forced to conclude that the scatter in rotation
rates for spots at any given latitude must be intrinsic to the spots
themselves.

Fig.~\ref{fig:resid} may provide a clue to the physical origin of this
scatter. Here we plot the residuals in rotational velocity
$$
v_{s}-v_{fit}=\frac{K R_{\star}}{v\sin i} 
\left(
\frac{2\pi}{P_{spot}} - \frac{2\pi}{P_{fit}} 
\right)
$$
measured relative to the mean fit, as a function of spot size.  The
scatter about the mean relation appears to be considerably greater for
the small spots than for the larger spots.  This size-dependent
scatter in the rotation rates of individual spots suggests that the
smaller spots may be more susceptible to random buffeting by
supergranular flows (see, e.g. \pcite{howard84}).  There does not,
however, appear to be any significant difference between the average
rotation rate of the large spots (i.e. those having $W>0.003$
km~s$^{-1}$), and that of the smaller spots.

\section{Conclusions}

The results of this investigation suggest that spectral subtraction,
least-squares deconvolution and matched-filter analysis provide a
powerful diagnostic method for identifying large numbers of individual
starspot signatures in trailed spectrograms of AB Doradus and, by
implication, other similar stars. The matched-filter analysis allows 
the positions and rotation periods of individual spots to be 
determined to high precision, and so provides a powerful alternative 
to methods involving regularised image reconstruction.

We have successfully measured the rotation periods of individual spots
at latitudes ranging from the stellar equator up to a latitude of
55$^{\circ}$.  The differential rotation law fitted to the larger
spots shows excellent agreement with the mean relation obtained by
\scite{donati97doppler}.  The rate of shear in the surface rotation
speed as a function of latitude appears to be very similar to the Sun,
with the equator pulling one full turn ahead of the pole every 110
days or so.

The rotation rates of the smaller spots at the low and intermediate
latitudes studied here appear to have a larger intrinsic
scatter than the large spots.  Further observations of similar or
better precision would be desirable to verify whether this size
dependence is present.  If confirmed, this result could give a first
glimpse of the turbulent sub-surface velocity field in a late-type
star other than the Sun.

\section*{ACKNOWLEDGMENTS}

We thank Keith Horne for helpful conversations and useful comments on
an early version of the manuscript.  An anonymous referee made several
useful suggestions for enhancing the clarity of the paper, and for
improving the treatment of limb darkening.  ACC acknowledges the
support of a PPARC Senior Research Fellowship during the course of
this work.  This paper is based on observations made using the 3.9-m
Anglo-Australian Telescope.  The project made use of support software
and data analysis facilities provided by the Starlink Project which is
run by CCLRC on behalf of PPARC.




\end{document}